\documentclass[twocolumn,showkeys,preprintnumbers,floatfix,amsmath,a4paper,nofootinbib,amssymb]{revtex4}

\usepackage[usenames,dvipsnames]{color}

\usepackage[latin1]{inputenc}
\usepackage{graphicx}
\usepackage{graphicx}
\usepackage{bm}
\usepackage{graphicx}

\begin{document}
\title{Typicality, entropy and the generalization of statistical mechanics}

\author{Bernat Corominas-Murtra}
\email{bernat.corominas-murtra@uni-graz.at}
\affiliation{Institute of Biology,  University Graz, Holteigasse 6, A-8010 Graz, Austria}

\author{Rudolf Hanel}
\email{rudolf.hanel@meduniwien.ac.at}
\affiliation{Complexity Science Hub Vienna, Josefst\"adter Strasse 39, 1080 Vienna, Austria}
\affiliation{SSCS, Medical University of Vienna, Spitalgasse 23, 1090 Vienna, Austria}

\author{$^*$Petr Jizba}
\email{p.jizba@fjfi.cvut.cz }
\affiliation{FNSPE,
Czech Technical University in Prague, B\v{r}ehov\'{a} 7, 115 19, Prague, Czech Republic}

%
%

\begin{abstract}
When at equilibrium, large-scale systems obey conventional thermodynamics because they belong to microscopic configurations (or states) that are {\em typical}. Crucially, the typical states usually represent only a small fraction of the total number of possible states, and yet the characterization of the set of typical states --- the {\em typical set} --- alone is sufficient to describe the macroscopic behavior of a given system. 
Consequently, the concept of typicality, and the associated {\em Asymptotic Equipartition Property} allow for a drastic reduction of the degrees of freedom needed for system's statistical description. The mathematical rationale for such a simplification in the description is due to the phenomenon of {\em concentration of measure}. The later emerges for equilibrium configurations thanks to very strict constraints on the underlying dynamics, such as weekly interacting and (almost) independent system constituents. 
The question naturally arises as to whether the concentration of measure and related typicality considerations can be extended and applied to more general complex systems, and if so, what mathematical structure can be expected in the ensuing generalized thermodynamics.
In this paper we illustrate the relevance of the concept of typicality in the toy model context of the ``thermalized'' coin and show how this leads naturally to Shannon entropy. We also show an intriguing connection: The characterization of typical sets in terms of R\'enyi and Tsallis entropies naturally leads to the free energy and partition function, respectively, and makes their relationship explicit.
Finally, we propose potential ways to generalize the concept of typicality to systems where the standard microscopic assumptions do not hold.
\end{abstract}

\maketitle


\section{Introduction}

A major conceptual contribution of statistical mechanics is that it successfully refocused its attention from phenomenological concepts such as heat and energy flow, which dominated thermodynamics in the second half of the 19th century, to the question of how the underlying microscopic dynamics occupies the space of potential configurations. This, in turn, made it possible to understand why a simple compact, macroscopic description is so efficient in describing microscopically diverse thermodynamic systems \cite{Gibbs:1902, Einstein:1902, khinchin:1949, Sklar:1993, Battermann:2001}. The key mathematical concept that is responsible for such a ``miraculous'' simplicity in the description is known as 
the concept of a {\em typical set} \cite{Cover:2012, Frigg:2009, Pitowsky:2012, Hanel:2023}. In fact, when at equilibrium, large-scale systems obey conventional thermodynamics because they lie in microscopic configurations (or states) that are typical. {\em Typical states} comprise a fraction of those possible states that carry total probability close to one. The set of these states is thus called the typical set.

Crucially, the set  of 
typical states comprises only a small fraction of the total number of possible states, and yet typical sets alone are sufficient to describe the macroscopic behavior of a given system. 
Consequently, the concept of typicality allows for a 
drastic reduction of the degrees of freedom needed for system's statistical description. A quantitative definition of typical states in weakly interacting systems is most easily provided by information theory~\cite{Cover:2012}. For continuous random variables the concept of typical sets is also studied in the framework of measure theory where it is tantamount to  the {\em concentration of measure} phenomenon~\cite{Talagrand:1995, Ledoux:05, Raginsky:2018}. The latter was popularized in the context of (multi)fractals by B.~Mandelbrot who also dubbed the phenomenon as {\em curdling}~\cite{Mand:83}. 
It is therefore the existence of the typical set of micro-states that allows the heat and energy flow considerations (underlying phenomenological thermodynamics) to be understood in terms of the occupancy of the state space in statistical mechanics~\cite{LEBOWITZ19931}. Furthermore, the equivalence between microcanonical and canonical ensemble description in the thermodynamic limit is, again, a direct consequence of the existence of a typical set of equal-energy microstates in the canonical ensemble~\cite{Nicholson:2016}. 

In Shannon's information theory, partitioning a set of states or sequences into those that are typical and those that are atypical
is possible due to the Asymptotic Equipartition Property (AEP) or Shannon--McMillan--Breiman theorem~\cite{Cover:2012}, which states that all the sequences in the typical set have almost the same probability to occur. There, the typical set and the AEP are instrumental in proving main results for the channel capacity and noiseless coding, as well as to provide a sound mathematical basis for various information compression strategies. Shannon's original proof of the AEP for independent and identically distributed (i.i.d.) sequence of random variables (as well as subsequent extensions to weakly dependent random variables) uses the weak law of large numbers. In such a context, Shannon entropy emerges as a natural tool for characterizing typical probabilities.
and, by extension, as a quantifier of the {\em cardinality} of the typical set. 
Moreover, the typical set contains sequences with a {\em sample entropy} (analog of Boltzmann entropy) that is close to the Gibbs--Shannon entropy. This is an information-theoretic equivalent of the celebrated Einstein's entropic principle~\cite{Einstein:1910}
(i.e., reversal of Boltzmann's entropic formula).
\begin{figure*}
\includegraphics[width=18.2cm]{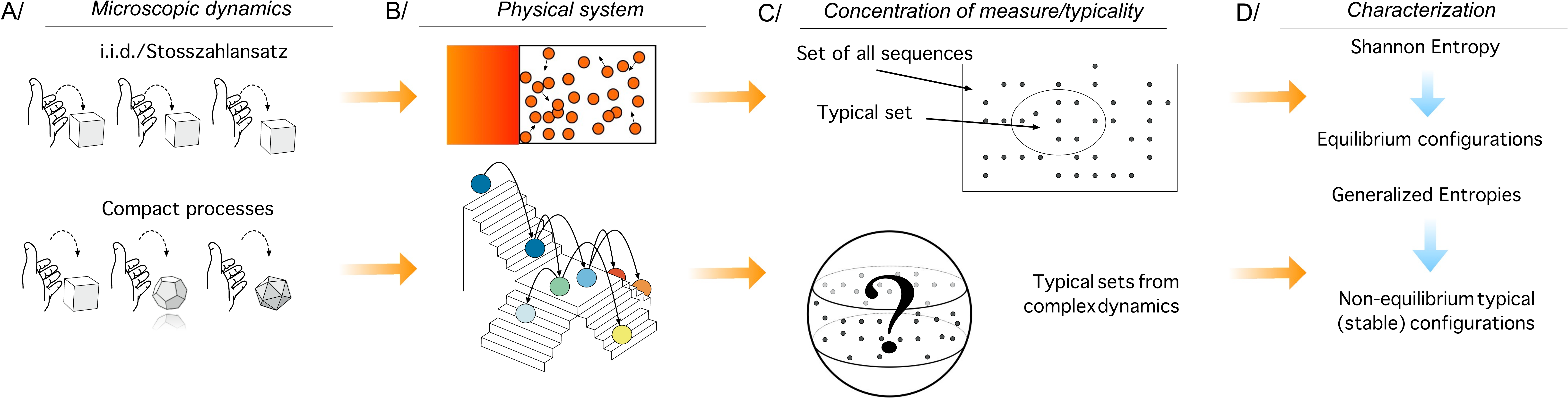}			%
\caption{A/ Different underlying microscopic hypothesis: (Top), the standard i.i.d. assumption, or its physical analogue, the Stosszahlansatz or chaos molecular hypothesis. (Bottom) a stochastic process whose sampling space grows in time, thereby violating the assumptions over which standard statistical mechanics is built. B/ (Top) A physical system in equilibrium whose microscopic dynamics obeys the molecular-chaos hypothesis and (Bottom) a toy representation of the collective behaviour of a physical system with increasing sampling space. C/ (Top) The standard assumptions of equilibrium statistical mechanics lead naturally to the concept of typical set. (Bottom) more complex dynamics may also have typical behaviours, albeit more complex to identify or characterize. D/ (Top) The existence of the typical set in equilibrium configurations gives rise to the Shannon entropy, as the natural functional accounting for the typical occupation of the sampling space. (Bottom) more complex dynamics still leading to typical behaviours may give rise to generalized forms of entropy and other functionals.}
\label{fig:AB}	
\end{figure*}

When the underlying dynamics satisfies appropriate boundary conditions, e.g. it is weekly interacting, then the conventional mathematical structure of equilibrium thermodynamics is directly implied by the concentration of measure phenomenon. This fact will be illustrated shortly with an example of a simple coin toss at different ``temperatures''.  Weakly interacting systems with (almost) independent system constituents  are epitomized, for instance, in the celebrated {\em Stosszahlansatz} hypothesis~\cite{Huang:1987} or Bogoliubov's no-correlation initial conditions~\cite{Bogoliubov:1962}. The question naturally arises as to whether the concentration of measure and related typicality can be applied to more general systems, such as complex dynamical systems, and if so, what mathematical structure can be expected in the ensuing generalized thermodynamics. The motivation for such a question is clear: the existence of typical behaviors implies a massive reduction of degrees of freedom and, in turn, triggers the emergence of macroscopic, interrelated functionals that allow us to characterize and define macroscopic observables. If typical sets and associated macroscopic functionals could be identified in systems with higher underlying microscopic complexity, this would open up the possibility of characterizing them (i.e. establishing predictive principles) in a way that would mimic equilibrium thermodynamics.
It is the aim of this paper to address this issue and put forward some of our preliminary results.

The paper is organized as follows.  In the next section, we analyze consequences of the concentration of measure and typicality in simple coin tossing systems. In particular, we discuss both microcanonical and canonical ensemble descriptions of such systems, and the role of temperature in the occupation of the state space. In Section~\ref{SEC.III}, we  extend our discussion beyond the conventional Shannon's paradigm, and we derive the conditions for typicality using both R\'enyi and Tsallis entropies. Interestingly, we find that the ensuing typical sets are well defined, allowing the occupation of the sampling space to be mapped into well-known functionals. In particular, typicality arising from R\'enyi entropies naturally involves the emergence of free energy, whereas for Tsallis entropies the typical set bounds are phrased in terms of the partition function.  We end with Section~\ref{sec:CSPs}, in which we briefly discuss a possible general framework in which the assumptions on the underlying dynamics are relaxed, leading to general forms of entropy characterizing the sampling space occupation even in cases with very complex dynamics and/or unstable sampling spaces. Conclusions and perspectives are finally summarized in Section~\ref{conclusions}. For the sake of clarity, some more technical considerations are relegated to appendix.

\section{Concentration of measure: tossing a coin}

In this section we provide the characterization of a fairly simple stochastic process ---  coin tossing. In particular, we show how the key observable quantities from information theory and statistical mechanics can be interpreted in terms of {\em measure concentration} and {\em typicality}, that is, in terms of state space occupation.

\subsection{A coin's ``microcanonical'' ensemble}
\label{sec:micro}

Suppose we have a system that can only be in two states $\{0,1\}$.
In the following, we will refer to this system as a  {\em coin}.
To start, let us assume that after running (or observing) our system $N$ times, it shows exactly $m_0$ $0$'s and $m_1$ $1$'s, with $N=m_0+m_1$.
This coin tossing process represents a two-valued discrete-time stochastic process known as a Bernoulli (binary) process.
Let $\sigma_{\mathrm g}(m_0,m_1)$ be a generic sequence of $1$'s and $0$'s, i.e.
\begin{eqnarray}
\sigma_{\mathrm g}(m_0,m_1) \ =  \ (1,0,0,1,1,0,. . . .,0,1)\, ,
\label{II.1.cc}
\end{eqnarray}
 containing exactly $m_0$ $0$'s and $m_1$ $1$'s. We refer to the set of all such sequences as $\Omega^N(m_0, m_1)$. The cardinality of such a set is given by the binomial coefficient:
 \begin{eqnarray}
 |\Omega^N (m_0,m_1)| = {N \choose m_0 }\quad.
 \end{eqnarray}
We note that all sequences (\ref{II.1.cc}) are equally likely. The {\em Boltzmann}-like entropy of the {\em macrostate} defined by the set $\Omega(m_0,m_1)$ is defined as
\begin{equation}
S(\Omega(m_0,m_1)) \ = \ \log |\Omega(m_0,m_1)|\, .
\label{eq:BoltzmannE}
\end{equation}
%
By using Stirling's approximation  
\begin{eqnarray}
\log(n!) \ = \ n \log n \ - \ n \ + \ \mathcal{O}(\log n)\, , 
\end{eqnarray}
which is valid for sufficiently large values of $n$, we might rewrite (\ref{eq:BoltzmannE}) as 
%
\begin{equation}
\log |\Omega(m_0,m_1)| \ \approx \  N H(\theta)\, .
\label{eq:NH}
\end{equation}
Here, the entropy per toss $H(\theta)$ is nothing but the {\em Shannon} entropy associated with the Bernoulli's random variable $\theta$, i.e. variable that acquires  two values ($0$ and $1$), with respective probabilities
\begin{eqnarray}
&&p(\theta = 0) \ \equiv  \ p(0) \ = \ \frac{m_0}{N}\, , \nonumber \\[2mm]
&&p(\theta = 1)\ \equiv \ p(1) \ = \ \frac{m_1}{N}\, ,
\end{eqnarray}
so that
\begin{equation}
H(\theta) \ = \ -\sum_{\theta \in\{0,1\}}p(\theta)\log p(\theta)\, .
\label{eq:Htheta}
\end{equation}
In passing we might note that  $H(\theta)$  is maximized by 
 a uniform distribution with a maximum $\log 2$, thus
\begin{equation}
H(\theta)\ \leq \  \log 2\, .
\label{eq:log2}
\end{equation}
One can interpret the above results as a toy representation of the {\em microcanonical} ensemble.

\subsection{A coin's ``canonical'' ensemble}
\label{sec:can}

Let us now consider the sequence of $N$ coin tosses for which we know only the {\em prior} probabilities of occurrence of $0$ or $1$, i.e. $p(0)$ or $p(1)$, respectively. 
Instead of a specific sequence of $0$'s and  $1$'s with fixed $m_0$ and $m_1$, we now have a i.i.d. sequence of random variables $\theta_1,\theta_2, . . .,\theta_N$ following the stationary distribution $\{p(0),p(1)\}$. We denote the state space of all possible sequences of length $N$ as $\Omega^N$, so that
\begin{eqnarray}
\Omega^N \ = \ \{0,1\}^N\, .
\end{eqnarray}
Clearly, the cardinality of this state space is
\begin{equation}
\left| \Omega^N\right|\ = \ 2^{N}\, .
\label{eq:OmegaN}
\end{equation}
Let us observe how the measure concentration phenomenon arises in this case as a consequence of the law of large numbers. Indeed, 
let us denote the average of the sum after $N$ trials as $\langle \theta_N\rangle$. Then for any $N$ we have 
\begin{eqnarray}
\langle \theta_N\rangle \ = \ Np(1)\, .
\end{eqnarray}
At this stage we might employ, e.g. Hoeffding's version of the law of large numbers (or Hoeffding's inequality)~\cite{Hoeffding:63}, which states that for all $\delta >0$
\begin{eqnarray}
&&\mathbb{P}\left\{\frac{1}{N}\left(\sum_{k=1}^N\theta_k- \langle \theta_N\rangle\right)\geq \delta\right\}\nonumber \\[2mm]
&& = \ \mathbb{P}\left\{\left(\frac{1}{N}\sum_{k=1}^N\theta_k-  p(1)\right)\geq \delta\right\}
\ \leq \ e^{-2\delta^2N}\, .~~~~~
\label{II.13.cf}
\end{eqnarray}
We see that most of the weight in long sequences is carried by sequences whose arithmetic average is close to $p(1)$, and that deviations from this behavior are extremely rare in long sequences. But can we know more? Apart from the inequality (\ref{II.13.cf}), it is important to know how the expected sequences look like. Obviously, not all $2^N$ sequences
of length $N$ have the same arithmetic mean. This leads to the concept of typicality and the associated concept of {\em typical set}.

\subsubsection{The typical set}\label{II.B.1.vb}
%
Let $p(\theta_1,\theta_2, \ldots,\theta_N)$
be the probability of observing the sequence $\theta_1,\theta_2, \ldots,\theta_N$ of i.i.d. random variables, then in the large $N$ limit the following form of AEP holds 
\begin{eqnarray}
-\frac{1}{N}\log p(\theta_1,\theta_2, \ldots,\theta_N)\ \to \ H(\theta)\, ,
\label{eq:1/Nlog}
\end{eqnarray}
where the convergence is in probability. This can be proved in various ways~\cite{Thomas:1988,Cover:2012}.
For instance, if we employ the fact that $\theta_k$'s are i.i.d.,  we have
\begin{eqnarray}
-\frac{1}{N}\log p(\theta_1,\theta_2, \ldots,\theta_N) \ = \ -\frac{1}{N} \sum_{k=1}^N \log p(\theta_k)\, ,
\end{eqnarray}
and by Hoeffding's inequality
\begin{eqnarray}
&&\mbox{\hspace{-8mm}}\mathbb{P}\left\{\left(-\frac{1}{N}\sum_{k=1}^N\log p(\theta_k) +  \langle \log p(\theta) \rangle\right)\geq \delta\right\}\nonumber \\[2mm]
&&=\ \mathbb{P}\left\{\left(-\frac{1}{N}\sum_{k=1}^N\log p(\theta_k) -   H(\theta) \right) \geq \delta\right\} \nonumber \\[2mm]
&&\leq \ e^{-2\delta^2N}\, .
\label{II.15.vg}
\end{eqnarray}
which directly implies (\ref{eq:1/Nlog}). The aforementioned AEP has important consequences for the understanding of the state space structure. In fact, let $A_\epsilon^N$ be the set of sequences of length $N$, $\sigma^N_1, \ldots \sigma^N_m,\ldots$ with a generic member $\sigma^N_{\mathrm g}$ satisfying
\begin{eqnarray}
e^{-N[H(\theta) +\epsilon]} \ \leq \ p(\sigma^N_{\mathrm g}) \ \leq \  e^{-N[H(\theta)-\epsilon]}\, , 
\label{II1.16.cf}
\end{eqnarray}
(for $\epsilon > 0$). For reasons to be seen shortly, the set $A_\epsilon^N$ is known as a typical set. 
Eq.~(\ref{II1.16.cf}) implies that 
\begin{eqnarray}
\left|-\frac{1}{N}\log p(\sigma^N_{\mathrm g}) - H(\theta)\right| \ \leq \ \epsilon\, ,
\end{eqnarray}
or in probability
\begin{eqnarray}
&&\mbox{\hspace{-12mm}}\mathbb{P}\left\{\left|-\frac{1}{N}\log p(\theta_1,\theta_2, \ldots,\theta_N) -   H(\theta) \right| < \epsilon\right\} \nonumber \\[2mm]
&=&\mathbb{P}\left\{ \theta_1,\theta_2, \ldots,\theta_N \in A_\epsilon^N\right\} \ > \ 1 - 2e^{-2\epsilon^2 N}\, .
\label{II.18.vb}
\end{eqnarray}
On the last line we used the inequality (\ref{II.15.vg}) with $\epsilon$ instead of $\delta$. Eq.~(\ref{II.18.vb})
shows that the probability of obtaining a sequence that belongs to the typical set $A_{\epsilon_N}^N$ converges  to one in the large $N$ limit. The result (\ref{II.18.vb}) directly implies that
\begin{eqnarray}
1 - 2e^{-2\epsilon^2 N} & < &  \mathbb{P}\left\{ \theta_1,\theta_2, \ldots,\theta_N \in A_\epsilon^N\right\} \nonumber \\[2mm]
&\leq & \sum_{\theta_1,\theta_2, \ldots,\theta_N \in A_\epsilon^N}  e^{-N[H(\theta)-\epsilon]}\nonumber \\[2mm]
&=&  \left|A_\epsilon^N\right| e^{-N[H(\theta) -\epsilon]}\, ,
\end{eqnarray}
where $\left|A_\epsilon^N\right| $ is the cardinality of
the set $A_\epsilon^N$. In the derivation we used
the defining relation (\ref{II1.16.cf}). 
Similarly, one can easily see (cf. e.g.~\cite{Cover:2012}) that
\begin{eqnarray}
1 &=& \sum_{\theta_1,\theta_2, \ldots,\theta_N \in \Omega^N} p(\theta_1,\theta_2, \ldots,\theta_N)\nonumber \\[2mm]
&\geq& \sum_{\theta_1,\theta_2, \ldots,\theta_N \in A_\epsilon^N} p(\theta_1,\theta_2, \ldots,\theta_N)\nonumber \\[2mm]
&\geq& \sum_{\theta_1,\theta_2, \ldots,\theta_N \in A_\epsilon^N} e^{-N[H(\theta) + \epsilon]}\nonumber \\[2mm]
&=&  \left|A_\epsilon^N\right| e^{-N[H(\theta) +\epsilon]}\, .
\end{eqnarray}
Thus, the cardinality of $A_\epsilon^N$ is constrained as follows
\begin{equation}
(1 - 2e^{-2\epsilon^2 N})e^{N[H(\theta) -\epsilon]} \ \leq \ \left|A_\epsilon^N\right| \ \leq \ e^{N[H(\theta) +\epsilon]}\, .
\label{eq:ineqTypi}
\end{equation}
%
%
%
%
%
%
From Eq.~(\ref{eq:OmegaN}) we get that the relative size of the typical set with respect to the cardinality of the state space $\Omega^N$ is bounded as 
\begin{eqnarray}
\frac{\left|A_{\epsilon}^N\right|}{\left|\Omega^N\right|} \ \leq \  e^{-N\delta}\,,
\end{eqnarray}
with $\delta = [\log 2 -H(\theta)] - \epsilon$. From Eq.~(\ref{eq:log2}) we can conclude that, outside the special case of equiprobability, the relative size of the typical set in relation to the state space decays at least exponentially fast with $N$, so
\begin{eqnarray}
\frac{\left|A_{\epsilon}^N\right|}{\left|\Omega^N\right|} \ \to \  0\, .
\label{23.cc}
\end{eqnarray}
This is a hallmark of typical sets, namely 
their relative cardinality is very small (absolute cardinality is often of measure zero), but they carry almost all of the probability.
In our case, we may notice that the measure becomes more and more {\em concentrated} around the tiny region of the state space where the sequences have (almost) constant probability $e^{-NH(\theta)}$. Consequently
for any $\sigma^N_{\mathrm g}\in A_{\epsilon}^N$
\begin{eqnarray}
p(\sigma^N_{\mathrm g})\ \sim \ e^{-NH(\theta)}\, ,
\end{eqnarray}
where the symbol $\sim$ denotes asymptotic equivalence to the first order in the exponent, that is $a\sim b$ if
\begin{eqnarray}
\frac{\log a}{\log b}\ \to \  1\, .
\end{eqnarray}
(if $b=1$, then $a \sim 1$ is satisfied by definition). Therefore, as the sequences get closer and closer to the typical ones, they tend to become equi-distributed. 


In passing we might observe that Eq.~(\ref{eq:ineqTypi}) allows to write the cardinality of the typical set  as
\begin{eqnarray}
\left|A_{\epsilon}^N\right| \ \sim \   e^{NH(\theta)}\, .
\end{eqnarray}
This could alternatively be rewritten as
\begin{eqnarray}
\frac{\log \left|A_{\epsilon}^N\right|}{N} \ \to \ H(\theta)\, . 
\end{eqnarray}
The latter means that the Shannon entropy is the logarithm of the cardinality of $\left|A_{\epsilon}^N\right|$  per particle. Interestingly, the cardinality of $\left|A_{\epsilon}^N\right|$ approaches in the large $N$ limit the cardinality of the set of sequences containing exactly $m_0=Np(0)$ {\em zeros} and $m_1=Np(1)$ {\em ones}, i.e. the cardinality of $\Omega(m_0,m_1)$ discussed in the previous subsection. 

%
%

From the foregoing coin tossing system  we can conclude that: a) the typical set concentrates around the subset of states from the state space that represent the ``microcanonical'' ensemble states from section~\ref{sec:micro},  b) 
the measure is distributed in such a way that, as $N$ increases, it gets closer and closer to the equi-distribution\footnote{Let us stress that the word {\em equi-distribution} implies that probabilities are equal {\em up to first order in the exponent}. }, i.e. the ``microcanonical'' ensemble distribution $1/|\Omega^N (m_0,m_1)| $. 
This latter property is typically referred to as 
AEP~\cite{Cover:2012}, and it is a particular case of the  measure concentration phenomenon. Finally, c) in the coin toss example, we have seen that entropy is just a (logarithmic) measure that quantifies the occupation of the state space and characterises the concentration of measure as long as $N$ is large. 

\subsubsection{ ``Temperature'' and occupation of the state space}
%
Let us now ``thermalize'' the above Bernoulli binary scheme. To this end, we consider the following scenario: First, we take the process described above as the {\em reference} process $\theta$, following $\{p(0), p(1)\}$, and formally associate a unit temperature with this process. Without loss of generality, we assume $p(0)> p(1)$ (the special  case when $p(0)= p(1) = 1/2$ will be discussed separately). Second, 
we deform the process $\theta$ with a single deformation parameter $\beta$ so that
\begin{eqnarray}
\left\{ p(0), p(1)\right\}\;\rightarrow\; \left\{ p_\beta(0), p_\beta(1)\right\}\, ,
\end{eqnarray}
where $p_\beta(k)$ is the {\em escort} transformation of the probability distribution, i.e.
\begin{equation}
p_\beta(k) \ = \ \frac{p^\beta(k)}{Z_\beta}\, ,
\label{eq:pbeta}
\end{equation}
%
with the normalization factor $Z_\beta$ defined as 
\begin{equation}
Z_\beta \ =  \ \sum_{k\in \{0,1\}}p^\beta(k)\, .
\label{eq_Zbetasum}
\end{equation}
We might note that
\begin{eqnarray}
\lim_{\beta\to\infty}p_\beta(0)\ = \ 1\, , \quad \lim_{\beta\to\infty}p_\beta(1) \ = \ 0\, ,
\end{eqnarray}
i.e., the process is ``frozen'' at high $\beta$'s, meaning that only a single result will materialise in repeated tosses. On the other hand
\begin{eqnarray}
\lim_{\beta\to 0}p_\beta(0) \ = \ \frac{1}{2}\, ,\quad \lim_{\beta\to 0}p_\beta(1) \ = \ \frac{1}{2}\, ,
\end{eqnarray}
i.e., at low $\beta$'s the process gets more and more close to a random fair coin. 
Note that if we had started with the reference process where 
$p(0) = p(1) = 1/2$, then the escort transformation would not change this distribution; in other words, the fair coin distribution is a {\em fixed point} of the escort transformation.   However, the latter fixed point is unstable, as any small deviation from the fair coin rule will cause the process to ``freeze'' in the $\beta \rightarrow \infty$ limit.
Consequently, $\beta$ behaves like the inverse temperature: The higher the temperature, the higher the randomness. In terms of state space occupation, it is easy to check that
\begin{eqnarray}
\lim_{\beta\to\infty} \left|A_{\epsilon}^N(\beta)\right| \ \sim \ 1\, ,
\label{2.33.cf}
\end{eqnarray}
that is, the effective size of the state space is reduced to a single state, the sequence $(0,0,\ldots ,0)$, in which the system is frozen. The latter state plays a role analogous to that of the pure state in quantum mechanics.  On the other hand
\begin{equation}
\lim_{\beta\to0} \left|A_{\epsilon}^N(\beta)\right|\ \ \sim \ \left|\Omega^N\right| \ = \ e^{N\log 2}\, ,
\label{eq:allfull}
\end{equation}
that is, the whole space of all possible binary sequences of length $N$. It is not difficult to see that for a generic $\beta$ we have
\begin{eqnarray}
\left|A_{\epsilon}^N(\beta)\right|\ \ \sim \ e^{NH(\theta(\beta))}\, ,
\label{2.35.cg}
\end{eqnarray}
where
\begin{eqnarray}
H(\theta(\beta)) \ = \ -\sum_{i\in\{0,1\}} p_\beta(i)\log p_\beta(i)\, ,
\label{II.36.hh}
\end{eqnarray}
is the Shannon entropy of the ``thermalized'' coin. It is not difficult to verify that
\begin{eqnarray}
\frac{d}{d\beta} H(\theta(\beta)) \ < \ 0\, ,
\end{eqnarray}
from which we can deduce that the cardinality of the typical set decreases monotonically with increasing $\beta$, and that the actual choice of the reference process is irrelevant for this type of behavior.  

Note that Eq.~(\ref{2.33.cf}) is valid for arbitrarily large but fixed $N$. 
It is interesting to know what happens when the limit $N \rightarrow \infty$ is perform first.
With the help of Eq.~(\ref{2.35.cg}), we can write 
\begin{eqnarray}
\lim_{\beta\to\infty} \lim_{N\to\infty}\frac{\log\left|A_{\epsilon}^N(\beta)\right|}{N} \ =  \ H(\theta(\infty)) \ = \ 0\, ,
\end{eqnarray}
which is reminiscent of the third law of thermodynamics, where in order to get correct entropy, one must first consider the thermodynamic limit and then the zero temperature limit~\cite{Lieb:1981}.

\section{Typicality in  coin tossing systems  --- going beyond Shannon's paradigm \label{SEC.III}}

From the preceding discussion, it follows that the concept of typical sets is closely related to the concept of Shannon entropy.  
It is thus natural to ask how unique is the role of Shannon entropy in determining typical sets.  To answer this, we will go
back to our $\theta$ process described by the distribution $\{p(0),p(1)\}$ and consider two important classes of non-Shannonian entropies, namely R\'enyi and Tsallis entropies.

The R\'enyi entropy of order $\alpha$ of the process $\theta$ is defined as~\cite{Renyi:1976a,JA}
\begin{equation}
H_\alpha(\theta) \ = \ \frac{1}{1-\alpha}\log\Big[\sum_{k\in \{0,1\}} p^\alpha(k)\Big]\, , 
\label{eq:RenyiDef}
\end{equation}
where $\alpha > 0$. By L'H\^opital's rule, the R\'enyi entropy converges to the Shannon entropy for $\alpha\to 1$, that is:
\begin{eqnarray}
\lim_{\alpha\to 1}H_\alpha(\theta) \ = \ H(\theta)\, .
\end{eqnarray}
Similarly, we may introduce the Tsallis entropy of order $\alpha$ as~\cite{Tsallis:1988a,Tsallis:book}
\begin{equation}
S_\alpha(\theta) \ = \ \frac{1}{\alpha-1}\Big[1- \sum_{k\in \{0,1\}} p^\alpha(k)\Big]\, ,
\label{eq:TsallisDef}
\end{equation}
where again 
\begin{eqnarray}
\lim_{\alpha\to 1}S_\alpha(\theta) \ = \ H(\theta)\, .
\end{eqnarray}
In this section we will explore how these two probability functionals characterize the concentration of measure in the Bernoulli binary scheme. In particular, we will see that the characterization of the typical set by the R\'enyi entropy and the Tsallis entropy leads in a natural way to the equilibrium free energy and the partition function, respectively. 


\subsection{Typical set from R\'enyi entropy} \label{Renyi}

Let us again consider a sequence of i.i.d. random variables $\theta_1, \ldots,\theta_N$ following the distribution $\{p(0),p(1)\}$ characterizing the Bernoulli process $\theta$. We can again use Hoeffding's inequality to show that 
%
%
\begin{eqnarray}
&&\mbox{\hspace{-10mm}}\frac{1}{1-\alpha}\log \left(\frac{1}{N}\sum_{k=1}^Np^{\alpha-1}(\theta_k)\right)
\ \to \  H_\alpha(\theta)\, ,
\label{eq:RenyLim}
\end{eqnarray}
where the convergence is understood as the convergence in probability.  Indeed, Eq.~(\ref{eq:RenyLim}) directly follows from Hoeffding's inequality
%
%
%
\begin{eqnarray}
&&\mbox{\hspace{-3mm}}\mathbb{P}
\left\{\left( \frac{1}{N}\sum_{k = 1}^N p^{\alpha-1}(\theta_k) - \langle p^{\alpha -1}(\theta) \rangle\right) 
\ge \delta\right\} \nonumber\\[2mm]
&& \le \  e^{-2\delta^2 N}\, ,
\label{44.cb}
\end{eqnarray}
where 
\begin{eqnarray}
\langle p^{\alpha -1}(\theta) \rangle &=& \frac{1}{N}\sum_{\theta_1, \ldots, \theta_N } p(\theta_1, \ldots, \theta_N) \sum_{k=1}^N p^{\alpha -1 }(\theta_k)\nonumber \\[2mm]
&=& \sum_{l \in \{0,1\}} p^{\alpha}(l)\, .
\end{eqnarray}
Similarly to Shannon entropy, we can associate with the expression~(\ref{eq:RenyLim}) a sequence of typical sets , which we will call {\em R\'enyi-type typical sets}.
In fact, let $B^N_{\epsilon}(\alpha)$ be the set of sequences of length $N$, i.e. $\sigma_1^N, \ldots, \sigma_m^N, \ldots$ with a generic member $\sigma_{\rm{g}}^N = (\theta_{{\rm{g}},1}, \ldots, \theta_{{\rm{g}},N})$ which satisfies
%
%
\begin{eqnarray}
N e^{(1-\alpha)H_\alpha(\theta)-\epsilon}&\leq& \sum_{k\leq N} p^{\alpha-1}(\theta_{{\rm{g}},k})\nonumber \\[2mm]&\leq& Ne^{(1-\alpha)H_\alpha(\theta)+\epsilon} \, ,
\label{46.hh}
\end{eqnarray}
(for arbitrary $\epsilon >0$). In Appendix~A, we use the concept of the Kolmogorov--Nagumo mean \cite{Nagumo:1930, Kolmogorov:1930} to show that this formulation of a typical set represents a natural generalization of the Shannon case.
For  $\epsilon \ll 1$ this can be equivalently rewritten as 
\begin{eqnarray}
\left| \frac{1}{N}\sum_{k=1}^N p^{\alpha-1}(\theta_{{\rm{g}},k})    - \langle p^{\alpha -1}(\theta) \rangle  \right| \ \le \ \tilde{\epsilon}\, ,
\label{47.kl}
\end{eqnarray}
where
\begin{eqnarray}
\tilde{\epsilon} \ = \  e^{(1-\alpha)H_\alpha(\theta)}\epsilon\, .
\end{eqnarray}
In probability we can write~(\ref{47.kl}) as  
\begin{eqnarray}
&&\mbox{\hspace{-7mm}}\mathbb{P}
\left\{ \left| \frac{1}{N}\sum_{k=1}^N p^{\alpha-1}(\theta_{k})    - \langle p^{\alpha -1}(\theta) \rangle  \right|  <  \tilde{\epsilon} \right\}\nonumber \\[2mm]
&&\mbox{\hspace{-5mm}} =\ 
\mathbb{P}\left(\theta_1, \theta_2, \ldots, \theta_N \in B^N_{\epsilon}(\alpha)\right) \ > \ 1 \ - \ 2e^{-2 \tilde{\epsilon}^2N}.~~~~
\label{48.nm}
\end{eqnarray}
On the last line we used~(\ref{44.cb}) and set $\tilde{\epsilon} = \delta$. 
From~(\ref{48.nm}) directly follows that
\begin{eqnarray}
\mathbb{P}\left(\theta_1, \theta_2, \ldots, \theta_N\in B^N_{\epsilon}(\alpha)\right)\ \to \  1\, .
\end{eqnarray}
So, similar to Shannon's case, in the large $N$ limit the set $B^N_{\epsilon}(\alpha)$ carries almost all probability --- justifying the name {\em typical set}. 

To find bounds on the cardinality of  $B^N_{\epsilon}(\alpha)$, we 
can follow the strategy from Section~\ref{II.B.1.vb}. In particular, to obtain the lower bound we can write (for $\alpha >1$)
\begin{eqnarray}
1 - 2e^{-2\tilde{\epsilon}^2 N} & < &  \mathbb{P}\left\{ \theta_1,\theta_2, \ldots,\theta_N \in B^N_{\epsilon}(\alpha)\right\} \nonumber \\[2mm]
&&\mbox{\hspace{-15mm}}= \ \sum_{\theta_1,\theta_2, \ldots,\theta_N \in B^N_{\epsilon}(\alpha)}  e^{\frac{N}{\alpha-1} \sum_{k\leq N} \frac{1}{N} \log(p^{\alpha-1}(\theta_k))}\nonumber \\[2mm]
&&\mbox{\hspace{-15mm}}\leq \ \sum_{\theta_1,\theta_2, \ldots,\theta_N \in B^N_{\epsilon}(\alpha)}  e^{\frac{N}{\alpha-1} \log\left[\sum_{k\leq N} \frac{1}{N} (p^{\alpha-1}(\theta_k))\right]} \nonumber \\[2mm]
&&\mbox{\hspace{-15mm}}= \ \left|B^N_{\epsilon}(\alpha)\right| e^{-N[H_{\alpha}(\theta) -\epsilon]}\, .
\end{eqnarray}
Here on the third line we used Jensen's inequality for concave functions (in this case logarithm) and on the last line we employed (\ref{46.hh}). Should we repeat the argument for $\alpha <1$, we would obtain
\begin{eqnarray}
1 - 2e^{-2\tilde{\epsilon}^2 N} & < &  \left|B^N_{\epsilon}(\alpha)\right| e^{-N[H_{2- \alpha}(\theta) -\epsilon]}\, .
\end{eqnarray}
As for the upper bound, we can write ($\alpha >1$)
\begin{eqnarray}
&&\mbox{\hspace{-10mm}}1 \ =\  \sum_{\theta_1,\theta_2, \ldots,\theta_N \in \Omega^N} p(\theta_1,\theta_2, \ldots,\theta_N)\nonumber \\[2mm]
&&\mbox{\hspace{-5mm}}\geq \  \sum_{\theta_1,\theta_2, \ldots,\theta_N \in B^N_{\epsilon}(\alpha)} p(\theta_1,\theta_2, \ldots,\theta_N)\nonumber \\[2mm]
&&\mbox{\hspace{-5mm}}=\ \sum_{\theta_1,\theta_2, \ldots,\theta_N \in B^N_{\epsilon}(\alpha)}  e^{\frac{N}{1-\alpha} \sum_{k\leq N} \frac{1}{N} \log(p^{1-\alpha}(\theta_k))}\nonumber \\[2mm]
&&\mbox{\hspace{-5mm}}\geq \ \sum_{\theta_1,\theta_2, \ldots,\theta_N \in B^N_{\epsilon}(\alpha)}  e^{\frac{N}{1-\alpha} \log\left[\sum_{k\leq N} \frac{1}{N} (p^{1-\alpha}(\theta_k))\right]} \nonumber \\[2mm]
&&\mbox{\hspace{-5mm}}= \  \left|B^N_{\epsilon}(\alpha)\right| e^{-N[H_{2-\alpha}(\theta) +\epsilon]}\, .
\label{53.cc}
\end{eqnarray}
Similarly for, $\alpha <1$ we get
\begin{eqnarray}
1 \ \geq \ \left|B^N_{\epsilon}(\alpha)\right| e^{-N[H_{\alpha}(\theta) -\epsilon]}\, .
\end{eqnarray}
Thus, the cardinality of $|B^N_{\epsilon}(\alpha)$ is constrained as follows ($\alpha > 1$)
\begin{eqnarray}
(1 - 2e^{-2\tilde{\epsilon}^2 N})e^{N[H_{\alpha}(\theta) -\epsilon]} &\leq& \left|B^N_{\epsilon}(\alpha)\right|\nonumber \\[2mm] &\leq&  e^{N[H_{2-\alpha}(\theta) +\epsilon]}
\, ,
\label{II.55.cv}
\end{eqnarray}
and similalry for $\alpha <1$.

Since maximum of R\'{e}nyi entropy for the Bernoulli binary scheme is $\log 2$, we get that
\begin{eqnarray}
\frac{\left|B^{N}_{\epsilon}(\alpha)\right|}{\left|\Omega^N\right|} \ \to \  0\, ,
\end{eqnarray}
which again shows that the relative cardinality decays at least exponentially with $N$.  In fact, one can prove even stronger statement~\cite{BHJ}, namely that the cardinality of the R\'{e}nyi-type typical set satisfies
\begin{eqnarray}
|B^{N}_{\epsilon}(\alpha)| \ \sim \ |A^{N}_{\epsilon/\alpha }|\, .
\end{eqnarray}
%

%
%
Now we rename $\alpha$ to $\beta$
and recall the definition of the partition function $Z_\beta$ from~(\ref{eq_Zbetasum}). 
%
%
Similarly as in the equilibrium thermodynamics, we can associate with  the partition function $Z_\beta$ the free-energy-like functional
\begin{equation}
F_\theta(\beta) \ = \ \log Z_\beta\, ,
\label{eq:freeenergy}
\end{equation}
which can be succinctly rewritten  as 
\begin{eqnarray}
F_\theta(\beta) \ = \ (1-\beta)H_\beta(\theta)\, .
\end{eqnarray}
In terms of free energy the typical set sequence identified through the R\'enyi entropy can be defined in a more compact way as the sequences $p_{\sigma_k^N}$ bounded as
\begin{eqnarray}
    e^{F_\theta(\beta) \ \!- \ \! \epsilon} &\leq& \frac{1}{N}\sum_{k\leq N} p^{\beta-1}(\theta_{{\rm{g}},k}) \nonumber \\[2mm]
    &\leq& e^{F_\theta(\beta) \ \! + \ \! \epsilon}\, .
    \label{eq:renyifree}
\end{eqnarray}
So, the typical set arising from the R\'enyi entropy gives rise to the free-energy-like functional. 

In passing we note that the Shannon entropy of the ``thermalized'' coin~(\ref{II.36.hh})  can be rewritten as 
\begin{eqnarray}
H(\theta(\beta)) &=& \beta H(p_\beta,p) \ + \ (1-\beta)H_\beta(\theta)\nonumber \\[2mm]
&=& \beta H(p_\beta,p) \ + \  F_\theta(\beta)\, .
\end{eqnarray}
Here the reference process $\theta$ has temperature $1$ and
\begin{eqnarray}
H(p_\beta,p)  \ = \ - \sum_{k \in \{0,1\} } p_{\beta}(k) \log p(k)\, ,
\end{eqnarray}
is an analogue of internal energy\footnote{Note that if one parametrizes $p(i)=e^{-u_i}$, where $u_i$ can be identified by some energy value, then $H(p_\beta,p)$ can be straightforwardly rewritten as $\langle u\rangle$. }.

This connection of the R\'enyi entropy with the free energy through a reference process was primarily reported in~\cite{Baez:2022} and studied in depth in \cite{Morales:2023}.

\subsection{Typical sets from Tsallis entropy} 

Now we turn to Tsallis entropy. Proceeding as above, one can straightforwardly proof that
\begin{equation}
\frac{1}{\alpha-1}\left(1- \sum_{k\leq N}\frac{1}{N}\ \! p^{\alpha-1}(\theta_k)\right) \ \to \  S_\alpha(\theta)\;.
\label{eq.RenyLim2}
\end{equation}
where the convergence is again meant in probability. The latter is a simple consequence of Eq.~(\ref{44.cb}). 

With the expression~(\ref{eq.RenyLim2}) we can associate typical sets that we will call {\em Tsallis-type typical sets}. In particular, let $C^N_{\epsilon}(\alpha)$ be the set of sequences of length $N$, i.e. $\sigma_1^N, \ldots, \sigma_m^N, \ldots$ with a generic member $\sigma_{\rm{g}}^N = (\theta_{{\rm{g}},1}, \ldots, \theta_{{\rm{g}},N})$ which satisfies
%
%
%
%
\begin{eqnarray}
&&N[1- (\alpha-1)S_\alpha(\theta)-\epsilon] \ \leq \ \sum_{k\leq N} p^{\alpha-1}(\theta_{{\rm{g}},k})\nonumber \\[2mm]&&\mbox{\hspace{25mm}}\leq \  N[1- (\alpha-1)S_\alpha(\theta)+\epsilon]\, .~~~~
\label{II.B.64.cb}
\end{eqnarray}
(for arbitrary $\epsilon >0$). The set $C^N_{\epsilon}(\alpha)$ is a typical set because
\begin{eqnarray}
\mathbb{P}\left(\theta_1, \theta_2, \ldots, \theta_N\in  C^N_{\epsilon}(\alpha)\right) \ \to \ 1\, .
\label{65.cv}
\end{eqnarray}
Before showing the validity of this relation, we will motivate the relation~(\ref{II.B.64.cb}). To this end, we rewrite Tsallis' entropy in terms of deformed logarithm as
\begin{eqnarray}
S_\alpha(\theta) \ = \  \sum_{k\leq N} p(\theta_k) \ln_{\alpha} \left(\frac{1}{p(\theta_k)} \right)\, ,
\end{eqnarray}
where the $\alpha$-logarithm is defined as
\begin{eqnarray}
\ln_{\alpha}(x) \ = \ \int_1^x dt \ \! t^{-\alpha} \ = \ \frac{1}{1-\alpha}\left(x^{1-\alpha } - 1 \right)\, . 
\end{eqnarray}
By rewriting Eq.~(\ref{II1.16.cf})  as
\begin{eqnarray}
H(\theta) +\epsilon &\geq&  \sum_{k \leq N} \frac{1}{N} \ \! \log \!\left(\frac{1}{p(\theta_{{\rm{g}},k})}  \right)\nonumber \\[2mm]
&\geq& H(\theta) +\epsilon\, ,
\end{eqnarray}
we might propose for the Tsallis-type typical set to satisfy the defining relation
\begin{eqnarray}
S_{\alpha}(\theta) + \epsilon &\geq&  \sum_{k \leq N} \frac{1}{N} \ \! \ln_{\alpha} \!\left(\frac{1}{p(\theta_{{\rm{g}},k})}  \right)\nonumber \\[2mm]
&\geq& S_{\alpha}(\theta) -\epsilon\, ,
\label{69.cc}
\end{eqnarray}
which indeed coincides with~(\ref{II.B.64.cb}). Here, the factor $1-\alpha$ was assimilated in the redefinition of $\epsilon$, so that the new $\epsilon$ is still positive. It is quite interesting to note that~(\ref{II.B.64.cb}) can also be rewritten in a form that is reminiscent of~(\ref{46.hh}), namely
\begin{eqnarray}
N \left[e_{\alpha}^{S_\alpha(\theta)-\epsilon}\right]^{1-\alpha}&\leq& \sum_{k\leq N} p^{\alpha-1}(\theta_{{\rm{g}},k})\nonumber \\[2mm]&\leq& N\left[e_{\alpha}^{S_\alpha(\theta)+\epsilon}\right]^{1-\alpha} \, ,
\label{70.cc}
\end{eqnarray}
where the $\alpha$-exponential is defined as~\cite{Tsallis:book}
\begin{eqnarray}
e_{\alpha}^x \ = \ [1 \ + \  (1-\alpha)x]_+^{{1}/{(1-\alpha)}}\, ,
\end{eqnarray}
with $[z]_+ = \max\{z, 0\}$. Inequality~(\ref{70.cc}) is valid for $\alpha <1$. Bounds must be reversed for $\alpha >1$.

Let us now turn back to~(\ref{65.cv}). In order to prove it, we rewrite~(\ref{II.B.64.cb}) as
\begin{eqnarray}
\epsilon  &\geq& \left| \frac{1}{N}\sum_{k=1}^N p^{\alpha-1}(\theta_{{\rm{g}},k})    - [ 1 + (1-\alpha)S_{\alpha}(\theta)]  \right| \nonumber \\[2mm]
&=& 
\left| \frac{1}{N}\sum_{k=1}^N p^{\alpha-1}(\theta_{{\rm{g}},k})    - \langle p^{\alpha -1}(\theta) \rangle  \right|\, ,
\label{72.cc}
\end{eqnarray}
which in probability can be written as 
\begin{eqnarray}
&&\mbox{\hspace{-7mm}}\mathbb{P}
\left\{ \left| \frac{1}{N}\sum_{k=1}^N p^{\alpha-1}(\theta_{k})    - \langle p^{\alpha -1}(\theta) \rangle  \right|  <  {\epsilon} \right\}\nonumber \\[2mm]
&&\mbox{\hspace{-5mm}} =\ 
\mathbb{P}\left(\theta_1, \theta_2, \ldots, \theta_N \in C^N_{\epsilon}(\alpha)\right) \ > \ 1 \ - \ 2e^{-2 {\epsilon}^2N}.~~~~
\label{48.nm}
\end{eqnarray}
The inequality is a consequence of~(\ref{44.cb}). This concludes our proof of~(\ref{65.cv}).

Let us now turn our attention to the cardinality of $C^N_{\epsilon}(\alpha)$.  Again, we can follow the strategy of section~\ref{II.B.1.vb}. In particular, to obtain the lower limit we can write (for $\alpha > 1$)
\begin{eqnarray}
1 - 2e^{-2{\epsilon}^2 N} & < &  \mathbb{P}\left\{ \theta_1,\theta_2, \ldots,\theta_N \in C^N_{\epsilon}(\alpha)\right\} \nonumber \\[2mm]
&&\mbox{\hspace{-15mm}} = \ \sum_{\theta_1,\theta_2, \ldots,\theta_N \in C^N_{\epsilon}(\alpha)}  e^{\frac{N}{\alpha-1} \sum_{k\leq N} \frac{1}{N} \log(p^{\alpha-1}(\theta_k))}\nonumber \\[2mm]
&&\mbox{\hspace{-15mm}}\leq \ \sum_{\theta_1,\theta_2, \ldots,\theta_N \in C^N_{\epsilon}(\alpha)}  e^{\frac{N}{\alpha-1} \log\left[\sum_{k\leq N} \frac{1}{N} (p^{\alpha-1}(\theta_k))\right]} \nonumber \\[2mm]
&&\mbox{\hspace{-15mm}}= \ \left|C^N_{\epsilon}(\alpha)\right| \left[e_{\alpha}^{S_{\alpha}(\theta) -\epsilon]}\right]^{-N}\, .
\end{eqnarray}
where on the last line we used~(\ref{72.cc}). For $\alpha <1$ we woud need to change $\alpha$ to $2-\alpha$. To obtain the upper bound we can write (fro $\alpha > 1$), cf.~(\ref{53.cc})
\begin{eqnarray}
&&\mbox{\hspace{-10mm}}1 \ =\  \sum_{\theta_1,\theta_2, \ldots,\theta_N \in \Omega^N} p(\theta_1,\theta_2, \ldots,\theta_N)\nonumber \\[2mm]
&&\mbox{\hspace{-5mm}}\geq \  \sum_{\theta_1,\theta_2, \ldots,\theta_N \in C^N_{\epsilon}(\alpha)} p(\theta_1,\theta_2, \ldots,\theta_N)\nonumber \\[2mm]
&&\mbox{\hspace{-5mm}}\geq \ \sum_{\theta_1,\theta_2, \ldots,\theta_N \in C^N_{\epsilon}(\alpha)}  e^{\frac{N}{1-\alpha} \log\left[\sum_{k\leq N} \frac{1}{N} (p^{1-\alpha}(\theta_k))\right]} \nonumber \\[2mm]
&&\mbox{\hspace{-5mm}}= \  \left|C^N_{\epsilon}(\alpha)\right| \left[e_{\alpha}^{S_{2-\alpha}(\theta) +\epsilon]}\right]^{-N}\, .
\end{eqnarray}
Consequently, the cardinality of $C^N_{\epsilon}(\alpha)$ is constrained as follows ($\alpha > 1$)
\begin{eqnarray}
(1 - 2e^{-2{\epsilon}^2 N})\left[e_{\alpha}^{S_{\alpha}(\theta) -\epsilon}\right]^N &\leq& \left|C^N_{\epsilon}(\alpha)\right|\nonumber \\[2mm] &\leq&  \left[e_{2-\alpha}^{S_{2-\alpha}(\theta) +\epsilon]}\right]^N
\, .
\label{II.55.cvv}
\end{eqnarray}
By identifying $\alpha$ with the ``inverse of the temperature'' $\beta$,
%
we obtain for the partition function that
\begin{eqnarray}
Z_\beta \ = \ 1-(\beta-1)S_\beta(\theta)\, .
\end{eqnarray}
As a result, the condition~(\ref{69.cc}) for typical Tsallis-type sequences can be rewritten in the form
\begin{eqnarray}
Z_\beta-\epsilon \ \leq \ \frac{1}{N}\sum_{k\leq N} p^{\beta-1}(\theta_{{\rm{g}},k}) 
\ \leq \ Z_\beta+\epsilon \, .
\end{eqnarray}
Therefore, the identification of the typical sequences from the Tsallis entropy leads to the emergence of the partition function of the ``thermalized process''.

\section{Entropy and typicality: going  beyond i.i.d.}
\label{sec:CSPs}

As we saw in the previous section, the {\em thermodynamic}-like structure of a process can be derived from the concentration of measure, under clearly stated underlying dynamical assumptions:  1) constant state space, i.e. no states appear or disappear through time, and 2) independence in the successive drawings, i.e. no time correlations are present. We then extended the notion of typicality and typical set by characterizing the occupation of the state space by different limiting procedure, giving rise to the R\'enyi and Tsallis entropies. In the latter context, we could observe that the corresponding entropies could not be formulated as {\em sample entropies}, i.e. logarithms (or deformed logarithms) of the cardinality of the ensuing typical sets --- as it was possible in the case of Shannon. On the other hand, the typical sets obtained were instrumental in defining ``equilibrium'' thermodynamic functions, namely free energy and partition function, without the need to introduce thermalized coins (and escort transformation).

Going beyond the simple structure of the Bernoulli binary
scheme (or, more generally, i.i.d. processes), the question arises: Could  a foregoing macroscopic picture emerge under more general assumptions?  For instance, for stochastic processes that may have growing or shrinking state spaces. Thus, the first task is to define a sufficiently general class of stochastic processes that includes the above cases and, if necessary, also processes violating the i.i.d. condition. We call such a class of processes {\em Compact Stochastic Processes} (CSP's)~\cite{Hanel:2023}. We are especially interested in finding a sample entropy of the system that is a functional of a trace-class type, and thus generalize the formula~(\ref{II1.16.cf}). In turn, this would represent the CSP's version of Einstein's celebrated entropic principle~\cite{Einstein:1910}. In the following, we will briefly outline a general strategy in this direction. More detailed discussion can be found in~\cite{Hanel:2023}.

\subsection{Basics: CSP's}

Let us consider a time-discrete stochastic processes $\eta$ \cite{Gardiner:1983, Feller:1991}. A realization of $N$ steps of the process is denoted as $\eta(N)$
\begin{eqnarray}
\eta(N) \ = \ \eta_1, \ldots,\eta_N\, ,
\end{eqnarray}
where $\eta_1, \ldots,\eta_N$ are random variables themselves. Note that, in different realizations of $t$ steps of the process, the sequence of random variables can be different, as the process may display path dependence, long term correlations, or changes of the phase space (either shrinking or expanding). We denote a particular trajectory (or a sample path) of the process as 
\begin{eqnarray}
x(t) \ \equiv \ x_1, \ldots ,x_N \ \in \  \Omega_\eta(N)\, .
\end{eqnarray}
Here $\Omega_\eta(N)$ is the set of all possible trajectories of the process $\eta$ after $N$ steps. We focus on the family of stochastic processes for which there exists i) a positive, strictly concave and strictly increasing function $\Lambda\in\mathcal{C}^2$ in the interval $[1,\infty)$, such that $\Lambda(1)=0$~\cite{Hanel:2011b, Hanel:2023}, and ii) a positive, strictly increasing function $g\in\mathcal{C}^2$, in the interval $(1,\infty)$, such that
\begin{eqnarray}
\lim_{N\to \infty}\ \! \frac{1}{g(N)}\Lambda\left(\frac{1}{p(\eta(N))}\right) \ = \ 1\, ,
\label{eq:toinftyeta}
\end{eqnarray}
in probability. Stochastic processes satisfying the above convergence relation are CSP's~\cite{Hanel:2023}. We recall that no assumptions were made about the process beyond the  convergence condition~(\ref{eq:toinftyeta}). In particular, we do not require independence of the successive values $\ldots,\eta_{N-1},\eta_N,\eta_{N+1} \ldots$ or stable state spaces from which the different elements of the sequence of random variables take values. We call the pair of functions $\Lambda, g$ {\em compact scale} of the process $\eta$, and note that a CSP can have more than one compact scale. 

The convergence condition (\ref{eq:toinftyeta}) implies the following asymptotic behavior for $\Lambda$
\begin{equation}
\lim_{z\to \infty}\frac{\Lambda(\lambda z)}{\Lambda(z)} \ = \ 1\, , \;\;\; \forall \lambda\in\mathbb{R}^+\, .
\label{eq:lambdaz}
\end{equation}
We refer to the set of functions $\Lambda$ as $\mathcal{L}$. Typical candidates for $\Lambda$ are of the form $\Lambda(z)=c\log^d(z)$, where $c,d$ are two positive, real valued constants or, more generally
\begin{eqnarray}
\Lambda(z) \ = \ c_1\log^{d_1}(1+c_2\log^{d_2}(1+ c_3\log^{d_3}(\ldots )))\, ,~~~~~
\end{eqnarray}
where $c_1, \ldots$ and $d_1, \ldots$ are positive, real valued constants\footnote{We observe that the Tsallis entropy would require an additional condition on the convergence, as $\Lambda\equiv \ln_\alpha$ alone does not satisfy the above condition directly.}. In previous approaches, these constants have been identified with scaling exponents that enable to classify the different potential growing dynamics of the phase space~\cite{Korbel:2018, Hanel:2023}. 

\subsection{The typical set and entropy in CSP's}

If, given a stochastic process $\eta$, there exists a compact scale $\Lambda,g$ for which the convergence condition (\ref{eq:toinftyeta}) 
holds, then
there exists a {\em typical set} $A_\epsilon^N\subseteq\Omega_\eta(N)$ of paths of the process $\eta$. The typical set will represent all probability in the limit of very large $N$, which means that although the potential set of paths can be arbitrarily large, only paths belonging to the typical set are expected to be effectively observed.   A path $x(N)=x_1 \ldots x_N$ of the process $\eta$ belongs to the typical set (or, alternatively, it is a {\em typical path}) if its associated probability to occur is bounded as
\begin{eqnarray}
\frac{1}{\Lambda^{-1}(g(N)(1+\epsilon))}&\leq& p(x(N))\nonumber \\[2mm]&\leq& \frac{1}{\Lambda^{-1}(g(N)(1-\epsilon))}\, ,
\end{eqnarray}
where $\Lambda^{-1}$ is the inverse of $\Lambda$, i.e., 
$(\Lambda^{-1}\circ \Lambda)(z)=z$, 
which exists given the assumption that $\Lambda$ is a monotonously increasing function.
In particular, one can prove~\cite{Hanel:2023} that for any $\epsilon>0$ there is a $N'>0$ such that, if $N>N'$
\begin{eqnarray}
\mathbb{P}\left(x(N)\in A_\epsilon^N \right) \ > \ 1-\epsilon\, .
\end{eqnarray}
In other words, the typical set usurps all probability, and the probability of observing non-typical paths becomes negligible. Note that since there is no assumption beyond the convergence condition~(\ref{eq:toinftyeta}), we cannot ensure the validity of tighter bounds on the concentration measure, as we did in section \ref{sec:can}, where we used Hoeffding's inequality, see e.g.~Eq.~(\ref{II.13.cf}). 
From the definition and properties of the typical set, it follows directly that there exists a non-increasing sequence of positive numbers $\epsilon_1, \ldots,\epsilon_N, \ldots$ such that $\epsilon_N\to 0$ (which we will write as $\epsilon_N \searrow 0$), defining a {\em sequence} of typical sets
\begin{eqnarray}
\ldots, A_{\epsilon_{t-1}}^{N-1}, A_{\epsilon_{t}}^{N}, A_{\epsilon_{t+1}}^{N+1}, \ldots\, .
\end{eqnarray}
by which:
\begin{equation} 
\mathbb{P}\left(x(N)\ \in \ A_{\epsilon_N}^N\right) \ \to  \ 1\, ,
\label{eq:PAto1}
\end{equation}
i.e., the typical set concentrates all the probability.

If, in the limit of large $N$, all contributions to the scaling factor $g(N)$ of paths outside the typical set vanish, one can rewrite the scaling factor as a trace-class entropic functional 
\begin{equation}
S_\Lambda(N) \ = \ \sum_{x(N) \ \in \ \Omega_\eta(N)}p(x(N))\Lambda\left(\frac{1}{p(x(N))}\right)\, ,
\label{eq:SLambdaTrace}
\end{equation}
which satisfies the first three of the four Shannon-Khinchin axioms for the entropic measure \cite{Hanel:2023}.
Consequently, the scaling term $g(N)$ can be identified with the {\em generalized} entropy $S_\Lambda(N)$ in simple CSP's. In turn, by construction, $S_\Lambda(N)$ converges to the (generalized) logarithm of the cardinality of the typical set, which describes the effective size of the state space. Indeed, for $\epsilon_N\searrow 0$
\begin{eqnarray}
\frac{\Lambda(|A_{\epsilon_N}^N|)}{S_\Lambda(N)} \ \to \ 1\, .
\end{eqnarray}
In addition, we observe that, for any $x(N)\in A_{\epsilon_N}^N$, with $\epsilon_N\searrow 0$, the following limit holds
\begin{equation}
\frac{\Lambda\left(\frac{1}{p(x(N))}\right)}{S_\Lambda(N)} \ \to \ 1\, .
\label{eq:Lambdapto1}
\end{equation}
This implies that the probabilities of the paths belonging to typical set are all equal upon the application of the generalized logarithm $\Lambda$. This might be viewed as a non-i.i.d generalization of the conventional AEP. 

With the above formalism, we therefore connected; 1) the microscopic dynamics of the system, 2) the effective increase of phase space (captured by the typical set evolution), and 3) a generalized entropic form $S_\Lambda$. 

\section{Discussion and Conclusions \label{conclusions}}

Characterizing the occupation of the state space is key to understanding the macroscopic properties of systems composed of many microscopic parts. The existence of the typical set, which is a direct consequence of the concentration of measure phenomenon, allows a massive reduction of degrees of freedom, giving rise to macroscopic functionals that characterize macroscopic configurations (i.e., macrostates). The typical set is thus the key concept that allows a rigorous justification of the reasoning behind statistical mechanics. Parallel to the considerations based on the concentration of measure, considerations on the deviations from the typical behaviors --- studied by the so-called {\em large deviations theory} --- provide very valuable information about the macroscopic behavior of the system~\cite{Touchette:2009}. In fact, the two are complementary and thus provide essential information for the possible thermodynamic interpretation of the different features of the sample space occupancy. 
The question remains whether and how the powerful concept of typicality and the resulting deviations from it can be extended to systems with more complex microscopic dynamics, such as non-i.i.d. systems.

To demonstrate the power of the unifying concept of typicality, we first provided a toy example, namely a ``thermalized'' coin, to illustrate how entropy, temperature, and occupancy of state space are interrelated. In this context we have shown that the typical set can be characterized not only by the Shannon entropy but also by the R\'enyi and Tsallis entropies. A remarkable observation here is that the characterization, considering different convergence criteria, using either R\'enyi or Tsallis-type typical sets, naturally leads to the free energy and partition function, respectively. So, in i.i.d. systems the typical set can be characterized from different entropic functionals, and this gives rise to the different thermodynamically relevant quantities. One might naturally expect that similar results will hold if, instead of i.i.d. sequences of random variables, we extend our considerations to weakly dependent random variables. This would naturally also cover a large portion of conventional equilibrium statistical thermodynamics. Beyond equilibrium, a general convergence condition can be postulated, namely, the CSP condition. 
Although CSP's encompass a broad class of microscopic stochastic dynamics, one can still define the state space occupation and the associated entropic functional. As we have seen in the case of the simple process of the coin toss, different approaches to the typical set may give rise to different, interrelated functionals representing different thermodynamically interpretable quantities. 
It remains an open question how different compact scales may relate (or even characterize) the same process, for general CSP's. 

In passing, it is interesting to mention a potential connection with the {\em coarse graining method} --- i.e. a concept from statistical mechanics introduced more than a century ago by Paul and Tanya Ehrenfest~\cite{Ehrenfest:1911} and further developed in 60's by Leo Kadanoff~\cite{Kadanoff}. The coarse-graining method procedure has proven to be a powerful procedure in statistical physics, especially when coupled with the concept of the renormalization group and the resulting portfolio of ideas and techniques that allow the systematic study of changes in a physical system as viewed on different length scales~\cite{Wilson:1975}. 
Coarse-graining --- or, more explicitly, the possibility of performing such an operation --- is often referred to as the key mechanism allowing for the massive collapse of degrees of freedom that leads to the thermodynamic interpretation of statistical ensembles. 
In this respect, one might argue that the existence of typical behaviors must underlie the success of the coarse-graining strategy to derive macroscopic behaviors from a deeper microscopic behavior. However, caution is required: In so-called renormalizable theories, the system at a coarse-grained scale will generally consist of self-similar copies of itself when viewed at a smaller coarse-grained scale, with different parameters describing the components of the system. The components, or fundamental variables, may relate to atoms, elementary particles, atomic spins, etc. No such changes in system's components are required when discussing typical sets.
There, the reduction in degrees of freedom is achieved purely as a result of the law of large numbers (or related isoperimetric inequalities), with all parameters of the macroscopic system remaining the same, regardless of whether one is working with the full state space or just a typical set.
The latter is certainly true for i.i.d. systems. For non-i.i.d. systems the dynamics and probability may be non-trivially intertwined --- with different system's parameters leading to different typical sets, and the renormalization group approach may then facilitate a new understanding of such a fact.

According to the open questions listed above, future work in this direction should address: 1) connections between Shannon, R\'enyi and Tsallis entropies, and uncover the relationship between state-space occupancy and thermodynamic (or information-theoretic) functionals --- even in the equilibrium case, where these relations are still poorly understood, 2)  possible connections between typicality, large deviation theory and the existence of coarse-grained descriptions of the system, and 3) the mathematical structure of a generalized statistical mechanics based on the existence of typicality in generic, more complex dynamics, e.g. in the framework of the CSP's. Crucially, these steps should go hand in hand with the derivation of empirically verifiable macroscopic observables.

\subsection*{Acknowledgments}

P.J. was in part supported by the Ministry of Education of the Czech Republic under grant M\v{S}MT RVO 14000. B.C.-M. acknowledges the support of the field of excellence ``Complexity of life in basic research and innovation'' of the University of Graz. 

\appendix

\section{Kolmogorov--Nagumo mean and R\'{e}nyi-type typical set }\label{secA1}

In this appendix we motivate the definition of the R\'{e}nyi-type typical set in equation ~(\ref{46.hh}). To do so, we note that the Kolmogorov--Nagumo (KN) mean of some random variable $X = \{x_1, \ldots, x_N\}$ with ensuing probability  $P = \{p(x_1), \ldots, p(x_N)$ is defined as
\begin{eqnarray}
\langle X \rangle_f  \ = \ f^{-1}\left( \sum _{k \leq N} p(x_k) f(x_k) \right) \, .
\end{eqnarray}
The KN average represents the most general class of averages compatible with Kolmogorov's probability postulates and subsumes the three conventional averages: arithmetic, geometric, and the harmonic mean. In particular, R\'{e}nyi's entropy can be formulated in terms of 
KN mean as
\begin{eqnarray}
H_{\alpha}(\theta) \ = \ f^{-1} \left( \sum _{k \leq N} p(\theta_k) f\Big(\log \frac{1}{p(\theta_k)}\Big)\right),
\label{A3.mn}
\end{eqnarray}
where the KN function 
\begin{eqnarray}
f(x) \ \equiv \ f_{\alpha}(x) \ = \ a \ \! e^{(1-\alpha) x} + b\, ,
\end{eqnarray}
with $a$ and $b$ being arbitrary (possibly $\alpha$-dependent) constants. If we choose $a = 1/(1-\alpha)$ and $b = -1/(1-\alpha)$, we obtain in the limit $\alpha \rightarrow 1$ the KN function $f_1(x) = x$, which yields Shannon's entropy. Note that two KN  functions which are linear functions of each other give the same mean and hence~(\ref{A3.mn}) is independent of actual values of $a$ and $b$. 

We note first that with the help of the KN mean, we can expect that similarly as
\begin{eqnarray}
&&-\frac{1}{N}\log p(\theta_1,\theta_2, \ldots,\theta_N)  \nonumber \\[2mm] &&= \   
- \sum_{k \leq N} \frac{1}{N} \log p(\theta_k)
\ \to \  H(\theta)\, ,
\label{eq:A1/Nlog}
\end{eqnarray}
also  the relation
\begin{eqnarray}
&&f^{-1}_{\alpha}\left(\sum_{k \leq N} \frac{1}{N} \ \! f_{\alpha}\Big(- \log p(\theta_k) \Big)\right) \nonumber \\[2mm] &&
= \ \frac{1}{1-\alpha}\log \left(\frac{1}{N}\sum_{k=1}^Np^{\alpha-1}(\theta_k)\right)
\to \ H_{\alpha}(\theta)\, , ~~~~~~
\end{eqnarray}
should hold. This was indeed proved in Sec.~\ref{Renyi}.

Let us now turn to Shannon's typical set $A_\epsilon^N$. This is defined  via the inequality
\begin{eqnarray}
e^{-N[H(\theta) +\epsilon]} \ \leq \ p(\sigma^N_{\mathrm g}) \ \leq \  e^{-N[H(\theta)-\epsilon]}\, , 
\label{A6.cc}
\end{eqnarray}
where $\sigma^N_{\mathrm g} = (\theta_{{\rm{g}},1}, \ldots, \theta_{{\rm{g}},N})$ is a general sequence from $A_\epsilon^N$. Eq.~(\ref{A6.cc}) can be rewritten equivalently as
\begin{eqnarray}
H(\theta) +\epsilon &\geq&  \sum_{k \leq N} \frac{1}{N} \Big(-\log p(\theta_{{\rm{g}},k})  \Big)\nonumber \\[2mm]
&\geq& H(\theta) -\epsilon\, .
\end{eqnarray}
Thus, for the R\'{e}nyi-type typical set, we might propose the defining relation
\begin{eqnarray}
H_{\alpha}(\theta) +\epsilon &\geq&  f^{-1}_{\alpha}\left(\sum_{k \leq N} \frac{1}{N} f_{\alpha}\Big(-\log p(\theta_{{\rm{g}},k})  \Big)\right)\nonumber \\[2mm]
&\geq& H_{\alpha}(\theta) +\epsilon\, .
\end{eqnarray}
This can be equivalently rewritten as
\begin{eqnarray}
N e^{(1-\alpha)H_\alpha(\theta)-\epsilon}&\leq& \sum_{k\leq N} p^{\alpha-1}(\theta_{{\rm{g}},k})\nonumber \\[2mm]&\leq& Ne^{(1-\alpha)H_\alpha(\theta)+\epsilon} \, ,
\end{eqnarray}
which coincides with (\ref{46.hh}).

\vspace{1cm}

\noindent{\bf{\large{Author contributions}}}

\vspace{3mm}

\noindent Conceptualisation (BC-M, RH, PJ), graphics (BC-M), 
calculations (BC-M, RH, PJ), and writing and editing (BC-M, RH, PJ).

\vspace{3mm}

\noindent {\bf{Data Availability Statement}} This manuscript has no
associated data or the data will not be deposited. [Author's
comment: The paper's contents are purely theoretical and
did not need any data].




\begin{thebibliography}{38}
\expandafter\ifx\csname natexlab\endcsname\relax\def\natexlab#1{#1}\fi
\expandafter\ifx\csname bibnamefont\endcsname\relax
  \def\bibnamefont#1{#1}\fi
\expandafter\ifx\csname bibfnamefont\endcsname\relax
  \def\bibfnamefont#1{#1}\fi
\expandafter\ifx\csname citenamefont\endcsname\relax
  \def\citenamefont#1{#1}\fi
\expandafter\ifx\csname url\endcsname\relax
  \def\url#1{\texttt{#1}}\fi
\expandafter\ifx\csname urlprefix\endcsname\relax\def\urlprefix{URL }\fi
\providecommand{\bibinfo}[2]{#2}
\providecommand{\eprint}[2][]{\url{#2}}

\bibitem[{\citenamefont{Gibbs}(2010)}]{Gibbs:1902}
\bibinfo{author}{\bibfnamefont{J.~W.} \bibnamefont{Gibbs}},
  \emph{\bibinfo{title}{Elementary Principles in Statistical Mechanics:
  Developed with Especial Reference to the Rational Foundation of
  Thermodynamics}}, Cambridge Library Collection - Mathematics
  (\bibinfo{publisher}{Cambridge University Press}, \bibinfo{year}{2010}).

\bibitem[{\citenamefont{Einstein}(1902)}]{Einstein:1902}
\bibinfo{author}{\bibfnamefont{A.}~\bibnamefont{Einstein}},
  \bibinfo{journal}{Annalen der Physik} \textbf{\bibinfo{volume}{314}},
  \bibinfo{pages}{417} (\bibinfo{year}{1902}).

\bibitem[{\citenamefont{Khinchin}(1949)}]{khinchin:1949}
\bibinfo{author}{\bibfnamefont{A.}~\bibnamefont{Khinchin}},
  \emph{\bibinfo{title}{Mathematical Foundations of Statistical Mechanics}},
  Dover Books on Mathematics (\bibinfo{publisher}{Dover Publications},
  \bibinfo{year}{1949}).

\bibitem[{\citenamefont{Sklar}(1993)}]{Sklar:1993}
\bibinfo{author}{\bibfnamefont{L.}~\bibnamefont{Sklar}},
  \emph{\bibinfo{title}{Physics and Chance: Philosophical Issues in the
  Foundations of Statistical Mechanics}} (\bibinfo{publisher}{Cambridge
  University Press}, \bibinfo{address}{New York}, \bibinfo{year}{1993}).

\bibitem[{\citenamefont{Battermann}(2001)}]{Battermann:2001}
\bibinfo{author}{\bibfnamefont{R.}~\bibnamefont{Battermann}},
  \emph{\bibinfo{title}{The Devil in the Details: Asymptotic Reasoning in
  Explanation, Reduction, and Emergence}} (\bibinfo{publisher}{Oxford
  University press: Oxford, UK}, \bibinfo{year}{2001}).

\bibitem[{\citenamefont{Cover and Thomas}(2012)}]{Cover:2012}
\bibinfo{author}{\bibfnamefont{T.~M.} \bibnamefont{Cover}} \bibnamefont{and}
  \bibinfo{author}{\bibfnamefont{J.~A.} \bibnamefont{Thomas}},
  \emph{\bibinfo{title}{Elements of information theory}}
  (\bibinfo{publisher}{John Wiley \& Sons:New York}, \bibinfo{year}{2012}).

\bibitem[{\citenamefont{Frigg}(2009)}]{Frigg:2009}
\bibinfo{author}{\bibfnamefont{R.}~\bibnamefont{Frigg}},
  \bibinfo{journal}{Philosophy of Science} \textbf{\bibinfo{volume}{76}},
  \bibinfo{pages}{997} (\bibinfo{year}{2009}).

\bibitem[{\citenamefont{Pitowsky}(2012)}]{Pitowsky:2012}
\bibinfo{author}{\bibfnamefont{I.}~\bibnamefont{Pitowsky}}, in
  \emph{\bibinfo{booktitle}{Probability in Physics}}, edited by
  \bibinfo{editor}{\bibfnamefont{Y.}~\bibnamefont{Ben-Menahem}}
  \bibnamefont{and} \bibinfo{editor}{\bibfnamefont{M.}~\bibnamefont{Hemmo}}
  (\bibinfo{publisher}{Springer: Berlin}, \bibinfo{year}{2012}), pp.
  \bibinfo{pages}{41--58}.

\bibitem[{\citenamefont{Hanel and Corominas-Murtra}(2023)}]{Hanel:2023}
\bibinfo{author}{\bibfnamefont{R.}~\bibnamefont{Hanel}} \bibnamefont{and}
  \bibinfo{author}{\bibfnamefont{B.}~\bibnamefont{Corominas-Murtra}},
  \bibinfo{journal}{Entropy} \textbf{\bibinfo{volume}{25}}
  (\bibinfo{year}{2023}).

\bibitem[{\citenamefont{Talagrand}(2001)}]{Talagrand:1995}
\bibinfo{author}{\bibfnamefont{M.}~\bibnamefont{Talagrand}},
  \bibinfo{journal}{Publications math\'{e}matiques de l'IH\'ES}
  \textbf{\bibinfo{volume}{81}} (\bibinfo{year}{2001}).

\bibitem[{\citenamefont{Ledoux}(2005)}]{Ledoux:05}
\bibinfo{author}{\bibfnamefont{M.}~\bibnamefont{Ledoux}},
  \emph{\bibinfo{title}{The Concentration of Measure Phenomenon}}
  (\bibinfo{publisher}{Providence, American Mathematical Society},
  \bibinfo{year}{2005}).

\bibitem[{\citenamefont{Raginsky and Sason}(2018)}]{Raginsky:2018}
\bibinfo{author}{\bibfnamefont{M.}~\bibnamefont{Raginsky}} \bibnamefont{and}
  \bibinfo{author}{\bibfnamefont{I.}~\bibnamefont{Sason}},
  \emph{\bibinfo{title}{Concentration of Measure Inequalities in Information
  Theory, Communications, and Coding: ThirdEdition}} (\bibinfo{publisher}{now
  publishers: Boston Delft}, \bibinfo{year}{2018}).

\bibitem[{\citenamefont{Mandelbrot}(1983)}]{Mand:83}
\bibinfo{author}{\bibfnamefont{B.}~\bibnamefont{Mandelbrot}},
  \emph{\bibinfo{title}{The fractal geometry of nature}}
  (\bibinfo{publisher}{Freeman, New York}, \bibinfo{year}{1983}).

\bibitem[{\citenamefont{Lebowitz}(1993)}]{LEBOWITZ19931}
\bibinfo{author}{\bibfnamefont{J.~L.} \bibnamefont{Lebowitz}},
  \bibinfo{journal}{Physica A: Statistical Mechanics and its Applications}
  \textbf{\bibinfo{volume}{194}}, \bibinfo{pages}{1} (\bibinfo{year}{1993}).

\bibitem[{\citenamefont{Nicholson et~al.}(2016)\citenamefont{Nicholson,
  Alaghemandi, and Green}}]{Nicholson:2016}
\bibinfo{author}{\bibfnamefont{S.~B.} \bibnamefont{Nicholson}},
  \bibinfo{author}{\bibfnamefont{M.}~\bibnamefont{Alaghemandi}},
  \bibnamefont{and} \bibinfo{author}{\bibfnamefont{J.~R.} \bibnamefont{Green}},
  \bibinfo{journal}{The Journal of Chemical Physics}
  \textbf{\bibinfo{volume}{145}}, \bibinfo{pages}{084112}
  (\bibinfo{year}{2016}).

\bibitem[{\citenamefont{Einstein}(1910)}]{Einstein:1910}
\bibinfo{author}{\bibfnamefont{A.}~\bibnamefont{Einstein}},
  \bibinfo{journal}{Annalen der Physik} \textbf{\bibinfo{volume}{33}},
  \bibinfo{pages}{1275} (\bibinfo{year}{1910}).

\bibitem[{\citenamefont{Huang}(1987)}]{Huang:1987}
\bibinfo{author}{\bibfnamefont{K.}~\bibnamefont{Huang}},
  \emph{\bibinfo{title}{Statistical Mechanics (Second ed.)}}
  (\bibinfo{publisher}{Wiley, New York}, \bibinfo{year}{1987}).

\bibitem[{\citenamefont{Bogoliubov}(1962)}]{Bogoliubov:1962}
\bibinfo{author}{\bibfnamefont{N.~N.} \bibnamefont{Bogoliubov}},
  \emph{\bibinfo{title}{{\rm in:} Studies in Statistical Mechanics, {\rm eds.
  I. J. de Boer and G. E. Uhleneck}}} (\bibinfo{publisher}{North-Holland,
  Amsterdam}, \bibinfo{year}{1962}).

\bibitem[{\citenamefont{Hoeffding}(1963)}]{Hoeffding:63}
\bibinfo{author}{\bibfnamefont{W.}~\bibnamefont{Hoeffding}},
  \bibinfo{journal}{Journal of the American Statistical Association}
  \textbf{\bibinfo{volume}{58}}, \bibinfo{pages}{13} (\bibinfo{year}{1963}).

\bibitem[{\citenamefont{Algoet and Cover}(1988)}]{Thomas:1988}
\bibinfo{author}{\bibfnamefont{P.~H.} \bibnamefont{Algoet}} \bibnamefont{and}
  \bibinfo{author}{\bibfnamefont{T.~M.} \bibnamefont{Cover}},
  \bibinfo{journal}{The Annals of Probability} \textbf{\bibinfo{volume}{16}},
  \bibinfo{pages}{899} (\bibinfo{year}{1988}).

\bibitem[{\citenamefont{Aizenman and Lieb}(1981)}]{Lieb:1981}
\bibinfo{author}{\bibfnamefont{M.}~\bibnamefont{Aizenman}} \bibnamefont{and}
  \bibinfo{author}{\bibfnamefont{E.~H.} \bibnamefont{Lieb}},
  \bibinfo{journal}{Journal of Statistical Physics}
  \textbf{\bibinfo{volume}{24}}, \bibinfo{pages}{279} (\bibinfo{year}{1981}).

\bibitem[{\citenamefont{R{\'{e}}nyi}(1976)}]{Renyi:1976a}
\bibinfo{author}{\bibfnamefont{A.}~\bibnamefont{R{\'{e}}nyi}},
  \emph{\bibinfo{title}{Selected Papers of Alfr\'{e}d R\'{e}nyi, 2nd Vol.}}
  (\bibinfo{publisher}{Akademia Kiado}, \bibinfo{address}{Budapest},
  \bibinfo{year}{1976}).

\bibitem[{\citenamefont{Jizba and Arimitsu}(2004)}]{JA}
\bibinfo{author}{\bibfnamefont{P.}~\bibnamefont{Jizba}} \bibnamefont{and}
  \bibinfo{author}{\bibfnamefont{T.}~\bibnamefont{Arimitsu}},
  \bibinfo{journal}{Annals of Physics} \textbf{\bibinfo{volume}{312}},
  \bibinfo{pages}{17} (\bibinfo{year}{2004}).

\bibitem[{\citenamefont{Tsallis}(1988)}]{Tsallis:1988a}
\bibinfo{author}{\bibfnamefont{C.}~\bibnamefont{Tsallis}}, \bibinfo{journal}{J.
  Stat. Phys.} \textbf{\bibinfo{volume}{52}}, \bibinfo{pages}{479}
  (\bibinfo{year}{1988}).

\bibitem[{\citenamefont{Tsallis}(2009)}]{Tsallis:book}
\bibinfo{author}{\bibfnamefont{C.}~\bibnamefont{Tsallis}},
  \emph{\bibinfo{title}{Introduction to Nonextensive Statistical Mechanics;
  Approaching a Complex World}} (\bibinfo{publisher}{Springer},
  \bibinfo{address}{New York}, \bibinfo{year}{2009}).

\bibitem[{\citenamefont{Nagumo}(1930)}]{Nagumo:1930}
\bibinfo{author}{\bibfnamefont{M.}~\bibnamefont{Nagumo}},
  \bibinfo{journal}{Japanese journal of mathematics :transactions and
  abstracts} \textbf{\bibinfo{volume}{7}}, \bibinfo{pages}{71}
  (\bibinfo{year}{1930}).

\bibitem[{\citenamefont{Kolmogorov}(1930)}]{Kolmogorov:1930}
\bibinfo{author}{\bibfnamefont{A.}~\bibnamefont{Kolmogorov}},
  \bibinfo{journal}{Accad. Naz. Lincei Mem. Cl. Sci. Fis. Mat. Natur. Sez.}
  \textbf{\bibinfo{volume}{12}}, \bibinfo{pages}{388} (\bibinfo{year}{1930}).

\bibitem[{\citenamefont{Corominas-Murtra
  et~al.}(2024)\citenamefont{Corominas-Murtra, Hannel, and Jizba}}]{BHJ}
\bibinfo{author}{\bibfnamefont{B.}~\bibnamefont{Corominas-Murtra}},
  \bibinfo{author}{\bibfnamefont{R.}~\bibnamefont{Hannel}}, \bibnamefont{and}
  \bibinfo{author}{\bibfnamefont{P.}~\bibnamefont{Jizba}}, \bibinfo{journal}{to
  be publised}  (\bibinfo{year}{2024}).

\bibitem[{\citenamefont{Baez}(2022)}]{Baez:2022}
\bibinfo{author}{\bibfnamefont{J.~C.} \bibnamefont{Baez}},
  \bibinfo{journal}{Entropy} \textbf{\bibinfo{volume}{24}}
  (\bibinfo{year}{2022}).

\bibitem[{\citenamefont{Morales et~al.}(2023)\citenamefont{Morales, Korbel, and
  Rosas}}]{Morales:2023}
\bibinfo{author}{\bibfnamefont{P.~A.} \bibnamefont{Morales}},
  \bibinfo{author}{\bibfnamefont{J.}~\bibnamefont{Korbel}}, \bibnamefont{and}
  \bibinfo{author}{\bibfnamefont{F.~E.} \bibnamefont{Rosas}},
  \bibinfo{journal}{New Journal of Physics} \textbf{\bibinfo{volume}{25}},
  \bibinfo{pages}{073011} (\bibinfo{year}{2023}).

\bibitem[{\citenamefont{Gardiner}(1983)}]{Gardiner:1983}
\bibinfo{author}{\bibfnamefont{C.~W.} \bibnamefont{Gardiner}},
  \emph{\bibinfo{title}{Handbook of Stochastic Methods for Physics, Chemistry
  and the Natural Sciences}} (\bibinfo{publisher}{Springer-Verlag: Berlin,
  Germany}, \bibinfo{year}{1983}).

\bibitem[{\citenamefont{Feller}(1991)}]{Feller:1991}
\bibinfo{author}{\bibfnamefont{W.}~\bibnamefont{Feller}},
  \emph{\bibinfo{title}{An Introduction to Probability Theory and Its
  Applications, Vol. 1,2}} (\bibinfo{publisher}{Wiley:New York, NY, USA},
  \bibinfo{year}{1991}).

\bibitem[{\citenamefont{Hanel et~al.}(2011)\citenamefont{Hanel, Thurner, and
  Gell-Mann}}]{Hanel:2011b}
\bibinfo{author}{\bibfnamefont{R.}~\bibnamefont{Hanel}},
  \bibinfo{author}{\bibfnamefont{S.}~\bibnamefont{Thurner}}, \bibnamefont{and}
  \bibinfo{author}{\bibfnamefont{M.}~\bibnamefont{Gell-Mann}},
  \bibinfo{journal}{Proceedings of the National Academy of Sciences USA}
  \textbf{\bibinfo{volume}{108}}, \bibinfo{pages}{6390} (\bibinfo{year}{2011}).

\bibitem[{\citenamefont{Korbel and Thurner}(2018)}]{Korbel:2018}
\bibinfo{author}{\bibfnamefont{R.}~\bibnamefont{Korbel},
  \bibfnamefont{J.~Hanel}} \bibnamefont{and}
  \bibinfo{author}{\bibfnamefont{S.}~\bibnamefont{Thurner}},
  \bibinfo{journal}{New Journal of Physics} \textbf{\bibinfo{volume}{20}},
  \bibinfo{pages}{093007} (\bibinfo{year}{2018}).

\bibitem[{\citenamefont{Touchette}(2009)}]{Touchette:2009}
\bibinfo{author}{\bibfnamefont{H.}~\bibnamefont{Touchette}},
  \bibinfo{journal}{Physics Reports} \textbf{\bibinfo{volume}{478}},
  \bibinfo{pages}{1} (\bibinfo{year}{2009}).

\bibitem[{\citenamefont{Ehrenfest and
  Ehrenfest-Afanasyeva}(1959)}]{Ehrenfest:1911}
\bibinfo{author}{\bibfnamefont{P.}~\bibnamefont{Ehrenfest}} \bibnamefont{and}
  \bibinfo{author}{\bibfnamefont{T.}~\bibnamefont{Ehrenfest-Afanasyeva}},
  \emph{\bibinfo{title}{The conceptual Foundations of the Statistical Approach
  in Mechanics}} (\bibinfo{publisher}{Cornell University Press},
  \bibinfo{year}{1959}).

\bibitem[{\citenamefont{Kadanoff}(1966)}]{Kadanoff}
\bibinfo{author}{\bibfnamefont{L.}~\bibnamefont{Kadanoff}},
  \bibinfo{journal}{Physics} \textbf{\bibinfo{volume}{2}}, \bibinfo{pages}{263}
  (\bibinfo{year}{1966}).

\bibitem[{\citenamefont{Wilson}(1975)}]{Wilson:1975}
\bibinfo{author}{\bibfnamefont{K.~G.} \bibnamefont{Wilson}},
  \bibinfo{journal}{Rev. Mod. Phys.} \textbf{\bibinfo{volume}{47}},
  \bibinfo{pages}{773} (\bibinfo{year}{1975}).

\end{thebibliography}
\end{document}